\documentclass[epj,twocolumn]{webofc}
\usepackage{graphicx}
\graphicspath{ {images/} }
\usepackage[rightcaption]{sidecap}
\usepackage{amsmath}
\usepackage{hyperref}
\usepackage{epstopdf}

\usepackage[varg]{txfonts}   
\wocname{MPGD,2015}
\woctitle{MPGD,2015}
\begin{document}
\title{Characterization of the GEM foil materials}

    
\author{ L.Benussi \inst{1}
\and S.Bianco \inst{1}
\and G.Saviano \inst{1,2}
\and S. Muhammad \inst{1,2,3,}
\thanks{Corresponding author; email address: Saleh.Muhammad@cern.ch}
\and D.Piccolo \inst{1}
\and G.Raffone \inst{1}
\and M.Caponero \inst{1}
\and L. Passamonti \inst{1}
\and D. Pierluigi \inst{1}
\and A. Russo \inst{1}
\and F. Primavera \inst{1}
\and S.Cerbelli \inst{2}
\and A. Lalli \inst{2}
\and M. Valente\inst{2}
\and M. Ferrini \inst{2}
\and B. Teissandier\inst{4}
\and M. Taborelli \inst{4}
\and M. Parvis \inst{5}
\and S. Grassini \inst{5}
\and J. Tirill$\acute{o}$ \inst{2}
\and F. Sarasini \inst{2}
\and A.V. Franchi \inst{2}
}



\institute{Laboratori Nazionali di Frascati - INFN, Frascati, Italy 
\and
University of Rome “La Sapienza” (IT) - Facoltà di Ingegneria, Ingegneria Chimica Materiali ed Ambiente, Italy
\and
 National Center for Physics, Quaid-i-Azam University Campus, Islamabad, Pakistan
\and
CERN- Vacuum, Surfaces and Coatings (VSC) TE Department, Switzerland
\and
Politecnico di Torino-Dipartimento di Fisica (DIFIS) Corso Duca degli Abruzzi, 24, I-10129 Torino, Italy}

\label{sec-2}
\abstract{

Systematic studies on the GEM foil material are performed to measure the moisture diffusion rate and saturation level. These studies are important because  the presence of this compound inside the detector's foil can possibly change its mechanical and electrical properties and, in such a way, the detector performance can be affected. To understand this phenomenon, a model is developed with COMSOL Multhiphysics v. 4.3 \cite{Comsol:4.3}, which described the 
adsorption and diffusion within the geometry of GEM foil, the concentration profiles and the time required to saturate the foil. The COMSOL model is verified by experimental observations on a GEM foil sample. This note will describe the model and its experimental verification results. 
}

%
\maketitle
\section{Introduction}
\label{intro}

To upgrade the forward muon region of the Compact Muons Solenoid (CMS) at Large Hadron Collider (LHC), the Gaseous Electron Multiplier (GEM) is considered a most suitable detector. Due to its high rate capability 1$MHz/cm^2$ \cite{Colaleo:2021453}, excellent position and timing resolution, the GEM detector can easily meet the requirements of the high pseudorapidity region where a high particles flux is expected. The major part of the detector is the GEM foil where charge amplification phenomena occur in the presence of intense electric filed in the conical shape micro holes within a thickness of 60$\mu$m (50$\mu$m kapton cladded with 10$\mu$m copper on both sides)
The installation of the triple-GEM detectors into muon stations at the end-cap region, as shown in figure 1, will improve muon momentum resolution, help to reduce the global muon trigger rate, assure a high muon reconstruction efficiency and increase offline muon identification coverage \cite{Abbaneo:2015nxa}.
The GEM foil is exposed to water because of production process, and because of possible leaks allowing atmospheric humidity to enter the gas volume. This study aims to measure the diffusion properties of GEM foils in order to obtain useful data to characterise the GEM foils mechanical tensile properties which are affected by humidity.

\begin{figure}
  \includegraphics[width=1.15\linewidth]{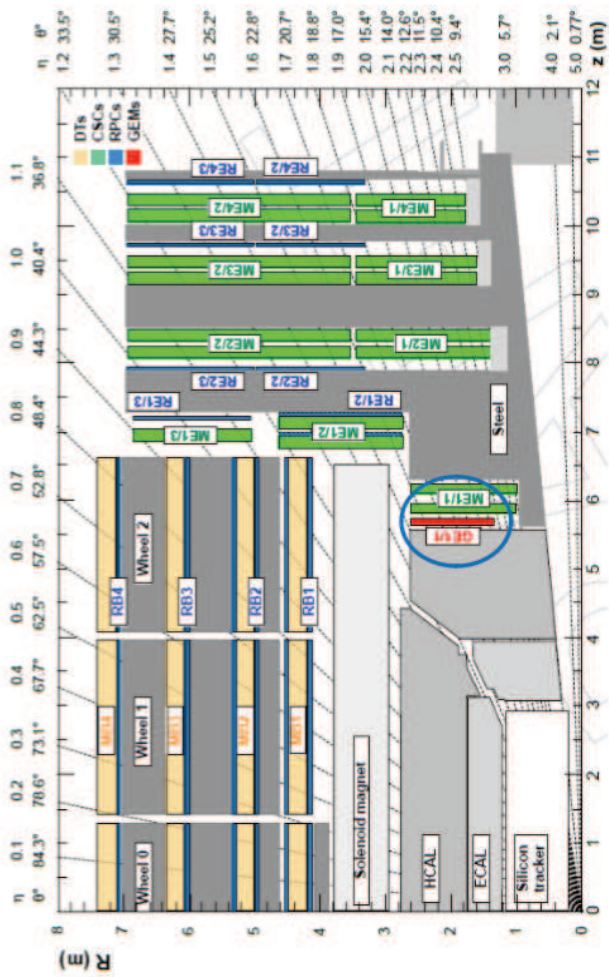}
  \caption{The GEM installation position at the CMS is highlighted in blue circle.}
\end{figure}

\section{Development of a diffusion model using COMSOL Multiphysics}
In the analysis of transport phenomena we considered the spread of $H_2O$ only from the holes surface where kapton is exposed to the environment. In fact  water cannot penetrate into the polymer from upper and lower surfaces due to copper layers which behave as a waterproof barrier.
The presence of any stagnant film around the surface of the hole is considered as negligible because the gas is continuously flushed inside the chamber, allowing to always have a constant and uniform concentration on the surface of the hole: this value represents the maximum concentration of water that can get inside the structure, in other words the value in full saturation conditions (after a theoretically infinite time).
In general the humidity that comes from the surrounding environment and spreads within the polymer, has condensation on the surface of solid and gas exchange (i.e. on the inner surface of the holes in the sheet). The relation that allows one to obtain the value of this concentration on the surface of a polyimide (kapton) is a function of the external humidity which is derived from the literature\cite{Sacher:1978}.

 \begin{equation}
  c1 = 4.520 * 10^{-4} * RH -8.319*10^{-4}        
 \end{equation}

where c1 is expressed in grams of water to grams of polymer. The formula used is empirical and it is compatible with the official data released by the manufacturer \cite{DUPONT:Kapton}, in case of RH = 100\% the maximum amount of water present in the polymer is equal to a 4.43\%, approximately consistent with  the 3\% declared by DuPont for Kapton. The above formula is not applicable for small values of relative humidity; considering, however, that the environment in which the detector will work does not provide RH > 4\%, it is possible to apply the previous formula while limiting to a 10\% error. The value of humidity chosen describes the unrealistic situation of 10\% of leaks inside the gas system of a  GEM detector.


The model developed takes into account the actual GEM foil geometry, including size and shape of the holes, for both the single and double mask process. Based on the problem symmetries, we selected the domain shown in figure 2, as representative periodic unit. This domain was discretized using triangular meshing.

\begin{figure}
  \includegraphics[width=1.15\linewidth]{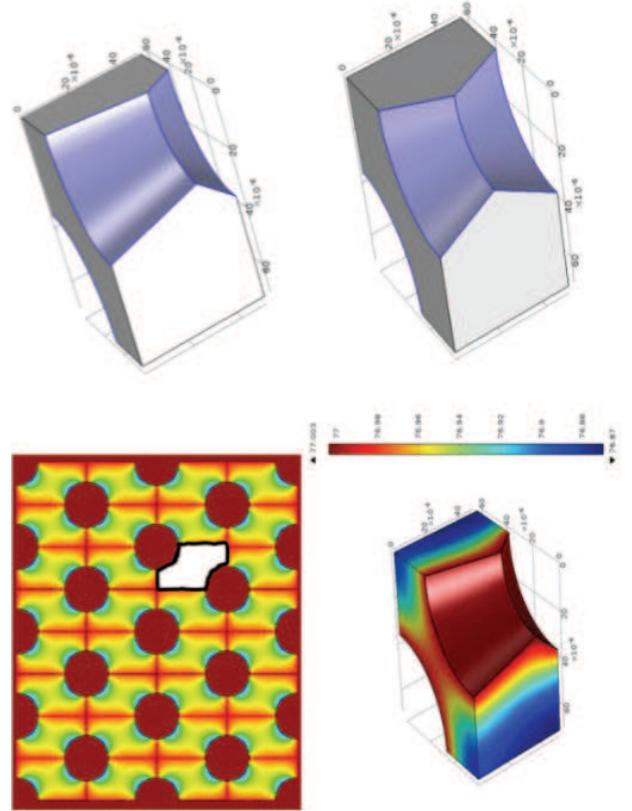}
  \caption{(top left)\cite{GEM:foil}. The selected area highlighted white (bottom left). The humidity absorption (top right). single mask (bottom right) double mask}
\end{figure}

The humidity exchange surfaces are colored in blue (figure 2), these being the only surfaces that allow the water to penetrate inside the foil structure. For all other faces coloured  in grey (figure 2)  it has set up the condition of no flow, this condition certainly valid for the upper and lower sides (where there is a copper coating).

The diffusion coefficient is taken from the previous laboratory tests, performed at CERN \cite{Diffusion:coefficent}, $ D=(2.11 \pm 0.18) 10^{-10}$ cm$^2$/s. The results obtained from the two geometries are very similar, the difference of the geometry is basically due to holes formations techniques which are called single and double masks techniques. 

The simulation results show that the saturation time is about 10 hours at RH =4\%, 14 hours at RH=40\% and 100\% as demonstrated in figures 3, 4, 5,  starting from a null value of water inside. The concentrations limits are different 75 mol/m$^3$, 1380 mol/m$^3$ and 3450 mol/m$^3$ at different relative humidity values 4\%, 40\% and 100\% respectively.  Along with RH value the saturation time also depends upon the geometry (size and shape) of holes and the value of the diffusion coefficient.


\begin{figure}
  \includegraphics[width=1.17\linewidth]{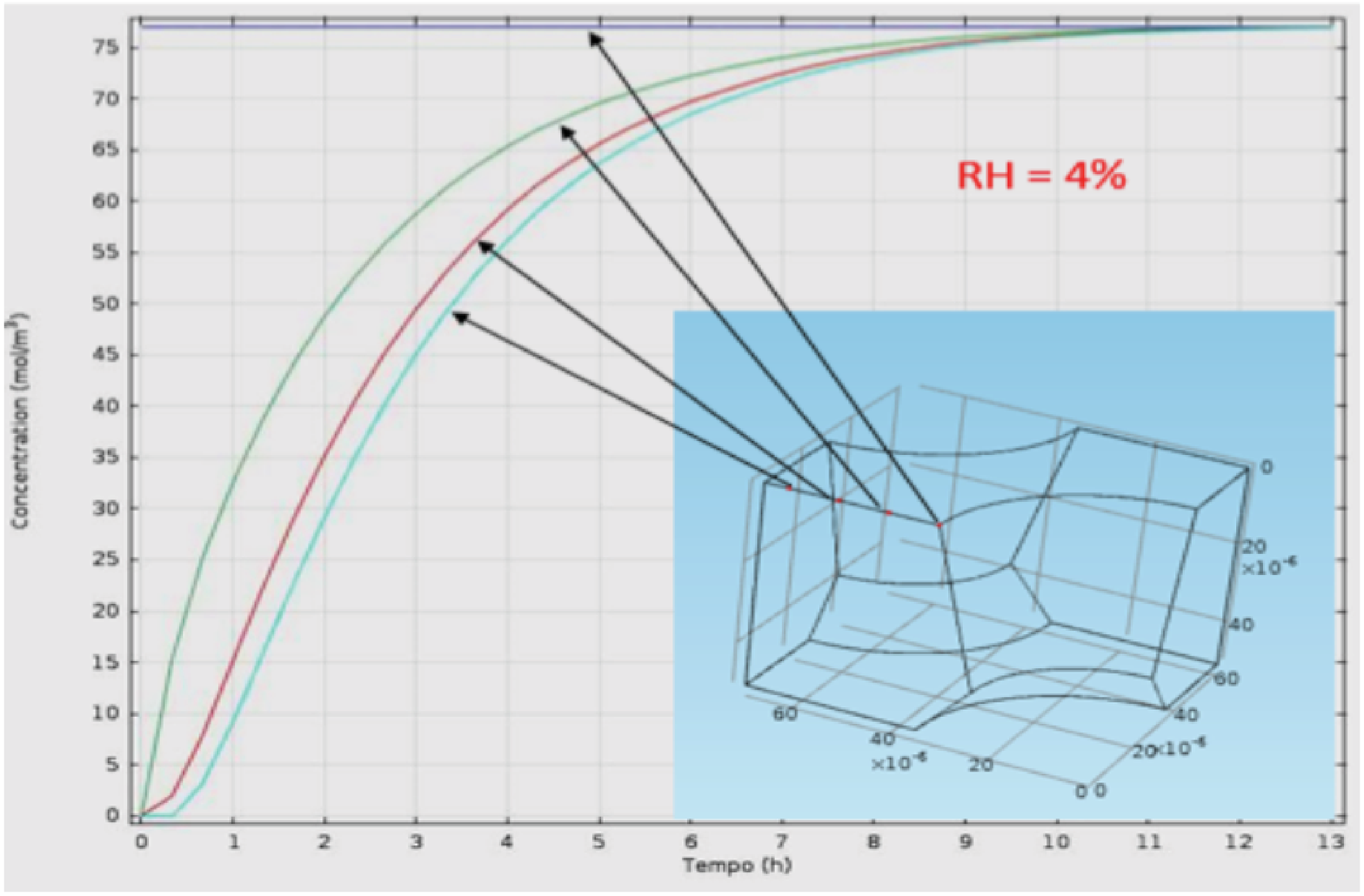}
  \caption{Time vs. concentration for different points of the geometry, with RH = 4\%}
  \includegraphics[width=1.22\linewidth]{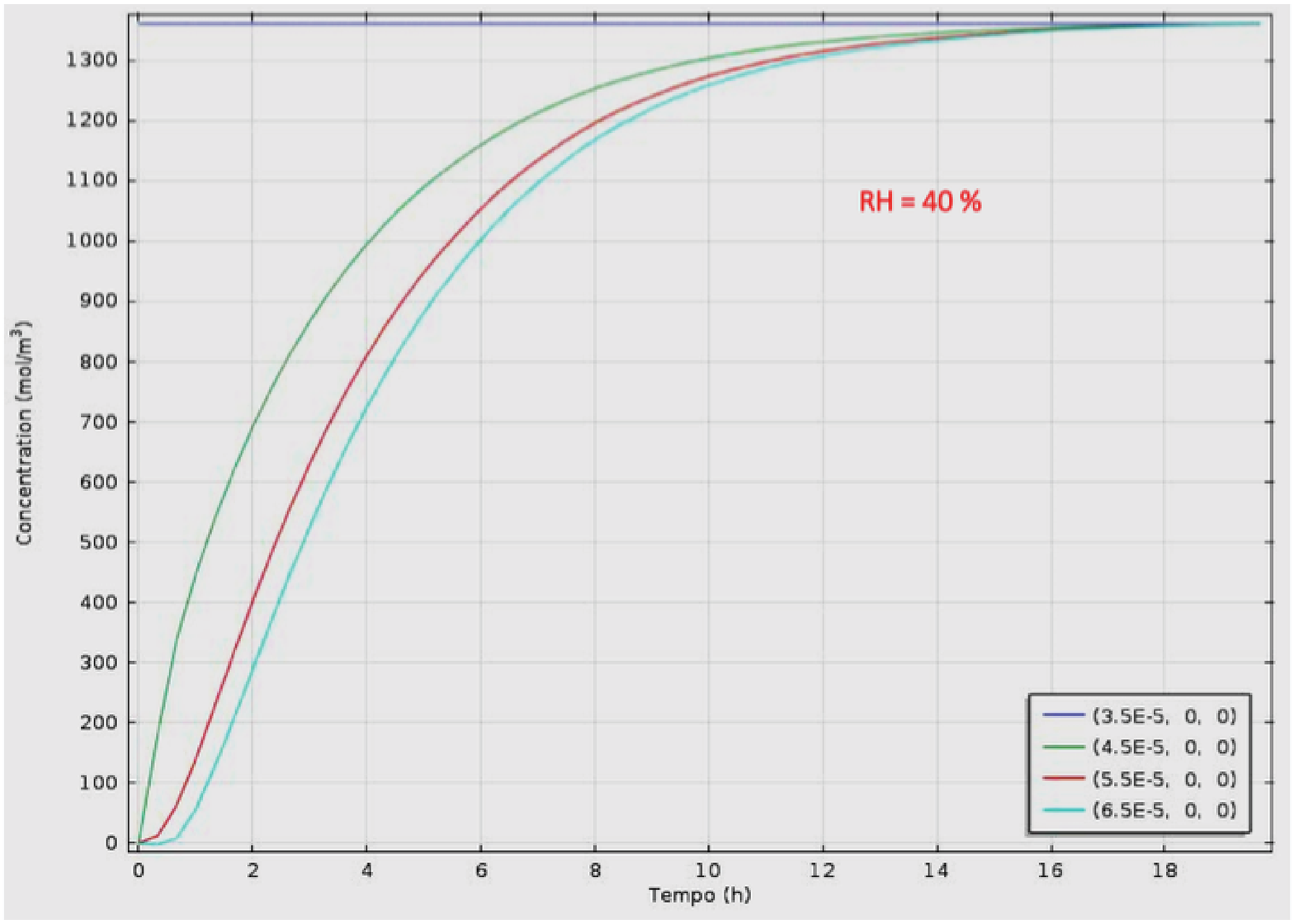}
  \caption{Time vs. concentration for the same points with RH = 40\%}
\end{figure}

\begin{figure}
  \includegraphics[width=1.25\linewidth]{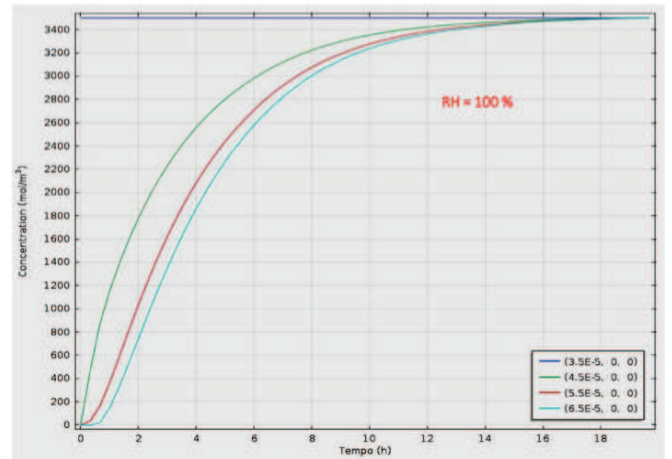}
  \caption{Time vs. concentration for different points of the geometry, with RH = 100\%}
\end{figure}


\section{Experimental verification of the diffusion model of water inside the GEM foil}
The diffusion model was verified to determine if it describes realistically the transport of water within the GEM structure.
To achieve this goal, a setup was built at the Frascati Laboratories of INFN to verify and improve the preliminary measurements  \cite{Diffusion:coefficent} performed at CERN,
particularly with respect to the weighing mode and the handling of systematic errors.

\subsection{Experimental setup}
The experimental setup, shown in figure 6, is composed of an analytical balance where the sample is supported by a wire. The analytical balance used for the measurements is a model Gibertini E42 S.The weight measurements are recorded by using a camera in front of the analytical balance display and connected to a computer figure 6 using a custom data acquisition program.

\begin{figure}
  \includegraphics[width=1.40\linewidth]{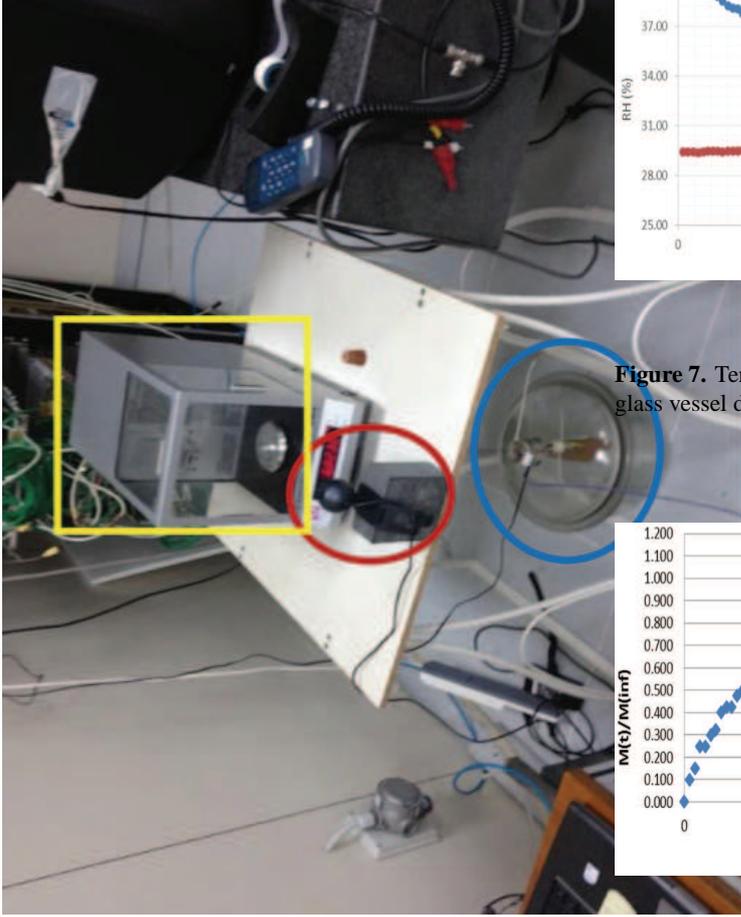}
  \caption{Data acquisition system (red) and instrumentation for monitoring environmental conditions parameters within the glass container (blue), weighing balance (yellow).}
\end{figure}

The environment conditions in which the GEM foil is exposed to water vapour are obtained by using the same type of vessel as used in the test held at the CERN laboratories \cite{Diffusion:coefficent};  however a mixture of salts and water inserted to condition uniformly the closed environment was used instead of a flux of humidified air. The mixture was composed of 170 ml of water and potassium carbonate, that allows to both saturate the system and to form  the residual deposit.

Sensors are located inside the glass vessel in order to record temperature and humidity with automatic data acquisition system. The room temperature and humidity where controlled and stabilized. The use of salts provided precise humidity inside the container thus helping reduce the systematic uncertainties during the measurements.
\begin{figure}
  \includegraphics[width=1.10\linewidth]{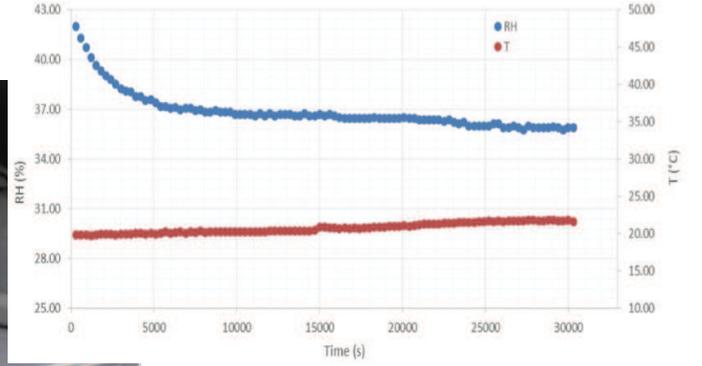}
  \caption{Temperature and relative humidity vs. time within the glass vessel during the test.}
\end{figure}
Oscillations where reduced by setting an additional weight on the balance plate. The precision reached a tenth of milligram.

\subsection{Test procedure}

A 105 mm * 50 mm * 60 $\mu$m sample of GEM foil is prepared and  conditioned in an oven at 105.0 $\pm$ 0.5 $^\circ C$ for 36 hours.  

 After the conditioning in oven, the sample is weighed and it is found a value of (0.6575 $\pm$ 0.0002) g.
 

The humidity and temperature range during the test remained RH = 37-38$\%$ and T = 19-20 $^\circ C$, shown in figure 7, the images are taken with every 5 minutes interval. The trends of the sample saturation are shown in figure 8. In 8-9 hours it is possible to completely saturate the GEM foil. The experimental results (figure 8) are comparable with the results obtained from the COMSOL model as described in figures 3,4,5.

\begin{figure}
  \includegraphics[width=1.0\linewidth]{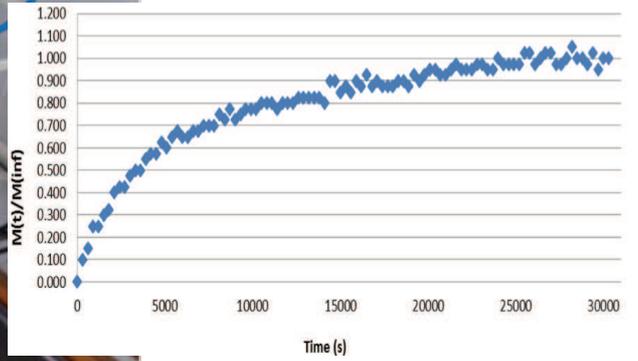}
  \caption{Humidity saturation trends in the GEM foil sample.}
\end{figure}

Furthermore, the measurements showed the presence of 0.004 g water at saturation level, compatible with what is obtained from the official datasheet by the manufacturer \cite{DUPONT:Kapton} or from the formula (1) for a value of RH = 37.5$\%$.

 \begin{equation}
  \acute{V}_{Kapton} = 0.2625 {\rm cm}^3
  \end{equation}
\begin{equation}
  V_{Kapton}=\acute{V}_{Kapton}*(1-\alpha) = 0.1835   {\rm cm}^3
\end{equation}
\begin{equation}
  P_{Kapton}=V_{Kapton}*\rho_{Kapton}= 0.2609 {\rm g}
\end{equation}
\begin{equation}
  P_{H_{2}O}=\left(\frac{c1}{ PM_{H_2O}} \right)*P_{Kapton}=0.0042 {\rm g}
\end{equation}

where  $\acute{V}_{Kapton}$ is the overall volume of the kapton in the sample, $V_{Kapton}$ is the volume of kapton multiplying with factor of holes which is $\alpha$ = 0.30 is the fraction of free space (due to the presence of the holes), $\rho_{Kapton}$=1.42 g/cm$^3$ is the density of kapton, $P_{Kapton}$ is the weights of kapton,  and $PM_{H_2O}$ is the molecular weight of water. Above calculations from equations 2 to 5, the mass of water in the kapton is calculated by using the experimetal conditions and data.

\section{Conclusions}
A model for describing the diffusion phenomenon within the GEM foil is developed. It is assumed that all the water coming from the humidity condenses on the surface of solid-gas exchange, i.e., the internal surface of the holes where the electronic avalanche will produce. This surface is the only available for the penetration of the water inside the foil, because the copper coating (on the top and bottom) behaves as a waterproof barrier. It is also assumed that near the surface of exchange there is not a stagnant film, hypothesis certainly validated by the presence of a gas flow in continuous circulation. Formula (1) is used to estimate the value of the concentration on the surface starting from a value of surrounding humidity; this value is assumed equal to 4$\%$ representing a hypothetical situation of 10$\%$ of leaks inside the gas system.

The results obtained from the COMSOL MULTIPHYSICS model showed that the time for having the complete saturation of a single mask GEM foil is about 10 to 14 hours depends upon humidity level.

In order to validate the COMSOL model results, a test is performed at the GEM foil sample, by doing continuous measurements of the weight in a well-conditioned system, built for the occasion: using a mixture of potassium carbonate and saturated water to get a stable value of RH in the range of 37-38$\%$ inside the container, the sample is hung to a wire, connected to the analytical  balance. After approximately 8 to 9 hours the equilibrium is achieved and the amount of water absorbed is 4 mg, this value being comparable with the official datasheet released by DuPont.

More studies are going on about the properties changes of the GEM foil due to the water absorption and  desorption from the system, in order to establish the standard conditions to obtain the longevity and the best performance of the GEM detector.



\end{document}